# Two-band effect on the vortex dynamics and critical current density in an anisotropic MgB$_2$ thin film

Jeehoon Kim,[1] N. Haberkorn,[2] E. Nazaretski,[3] R. de Paula,[1] X. X. Xi,[4] T. Tajima,[1] R. Movshovich[1] and L. Civale.[1]

[1]*Los Alamos National Laboratory, Los Alamos, NM 87545 USA.*

[2]*Centro Atómico Bariloche, Bariloche, 8400, Argentina.*

[3]*Brookhaven National Laboratory, Upton, NY 11973, USA.*

[4]*Department of Physics, Temple University, Philadelphia, PA 19122, USA.*

We report the influence of intrinsic superconducting parameters on the vortex dynamics and the critical current densities of a MgB$_2$ thin film. The small magnetic penetration depth of $\lambda$ = 50 nm at $T$=4 K is related to a clean $\pi$-band, and transport and magnetization data show an upper critical field similar to those reported in clean single crystals. We find a high self-field critical current density $J_c$, which is strongly reduced with applied magnetic field, and attribute this to suppression of the superconductivity in the $\pi$-band. The temperature dependence of the creep rate $S(T)$ at low magnetic field can be explained by a simple Anderson-Kim mechanism. The system shows high pinning energies at low field that are strongly suppressed by high field, which is consistent with a two-band contribution.

The binary MgB$_2$ compound has been subject of intense studies since the discovery of its superconductivity.[1] The nature of a two band superconductivity and the properties of interest to technological applications motived both basic and applied research.[2,3,4] Recently, research on MgB$_2$ has been revitalized in the context of iron-based superconductors, because MgB$_2$ offers a potential venue for understanding multiband superconductivity in pnictides, due to its relatively simple two-band structure. The MgB$_2$ system presents two distinct *s*-wave superconducting gaps - from the two dimensional (2D) $\sigma$ band and the three dimensional (3D) $\pi$ band, with superconducting energy gaps of $\Delta_\sigma(0) \approx 7.2$ mV and $\Delta_\pi(0) \approx 2.3$ mV, which leads to both inter- and intra-band scattering, and results in a rich array of superconducting properties.[5,6,7] The strong suppression of the π-band superconductivity with increasing magnetic field in clean systems can be inferred from differences between the upper critical field anisotropy ($\gamma_{Hc2} = H_{c2}^{ab}/H_{c2}^{c} = \xi_{ab}/\xi_{c}$)[8] and the magnetic penetration depth anisotropy ($\gamma_\lambda = \lambda_{ab}/\lambda_c$).[9,10] The difference of the magnetic field dependence of the two gaps is also manifested in the Ginzburg-Landau parameter $k = \lambda/\xi$, which ranges from $k$=2–3 at low field to $k$=7 close to $H_{c2}$.[11] In dirty samples disorder (via several mechanisms) affects the intra-band diffusivity in each band, and thereby the resulting physical properties.[6] Recently we reported the influence of large intra-band diffusivity in the 3D π band on the superconducting properties in a MgB$_2$ thin film in the dirty limit,[12] which results in larger λ and smaller $\gamma_{Hc2}$ than those found in clean single crystals.[11]

The unconventional superconducting properties in MgB$_2$ allow for a study of the influence of two superconducting gaps on the vortex dynamics, a phenomenon relevant to the physics of pnictides. The strong suppression of the superconductivity in the π band with magnetic field $H$ (below 1 T), is directly related to the effect of the changing λ on the depairing critical current ($J_0$),[11] and should influence $J_c(H)$ in a way that goes beyond the type of pinning centers alone. In general, in the context of defects as pining centers, clean single crystals show low $J_c$ values,[13,14] while thin films show higher $J_c$ values.[15,16,17] The vortex dynamics in dirty MgB$_2$ films is characterized by low creep rates ($S$) and high pinning energies,[18] a behavior intermediate between the low $T_c$ superconductors and the high $T_c$ cuprates.[19] In addition, very high $J_c$ values have been reported in clean MgB$_2$ films at low $H$,[4] indicating that the nature of pinning in MgB$_2$ films should be considered in combination of both intrinsic superconducting properties and pinning landscape.

In this work we present the intrinsic superconducting properties and their influence on the vortex dynamics in a MgB$_2$ thin film. Our results show that the film is in the clean limit, and at low $H$ the superconducting properties are dominated by a clean π− band. At low $T$ and low $H$, the film exhibits high $J_c$ values. However, the superconducting properties are affected significantly with increasing $H$, which can be associated with the magnetic field suppression of superconductivity in the clean π− band, resulting in a drastic change of the intrinsic superconducting properties of the material.

**Experimental**

The MgB$_2$ thin film with the thickness of 310 nm was grown on a *c*-cut sapphire by the hybrid physical chemical vapor deposition (HPCVD) technique. A detailed description of the epitaxial growth of MgB$_2$ by HPCVD has been reported elsewhere.[20] The phase purity for each crystal was examined by x-ray diffraction (XRD). The *T* and **H** dependence of the magnetization was studied using a superconductor quantum interference device (SQUID) magnetometer. A direct penetration depth ($\lambda$) measurement at 4 K was performed by magnetic force microscopy (MFM) based on a direct comparison of the Meisnner response forces between the sample and a Nb reference *in situ*.[21] The critical currents densities ($J_c$) were estimated applying the Bean critical-state model to the magnetization data, obtained from hysteresis loops, $J_c = \dfrac{20\Delta M}{tw^2(l - w/3)}$, where $\Delta M$ is the difference in magnetization between the top and bottom branches of the hysteresis loop, and *t*, w, and *l* are the thickness, width, and length of the sample ($l > w$), respectively. The creep measurement [$J_c(t)$] was recorded over a time period of one hour. The initial time was adjusted considering the best correlation factor in the log-log fitting of the $J_c(t)$ dependence. The initial critical state for each creep measurement was prepared by applying a field of $H \sim 4\,H^*$, where $H^*$ is the field for the full-flux penetration.[22] Electrical resistivity was measured using the standard four-probe technique. The samples were mounted in a rotatable probe and the measurements were performed in applied magnetic fields of between 0 and 9 T. The angular dependence $J_c(\theta)$ was measured from the current – voltage (I-V) curve by using the 1 µV criteria. Transport measurements were conducted with applied current (**J**) perpendicular to **H** in a maximum Lorentz force configuration. The angle θ is defined between the applied field and the *c*-axis of the MgB$_2$ (perpendicular to the surface).

**Results and discussion**

The superconducting critical temperature ($T_c$) and its transition width are $T_c$ =39.7 K and $\Delta T_c$ = 0.1 K, respectively. The residual resistance ratio ($\rho^{300\,K}/\rho^{42\,K}$) is ≈ 18. Figure 1(a) shows the temperature dependence of the upper critical field ($H_{c2}$) and the irreversibility line ($H_{irr}$) with the magnetic field (**H**) parallel (//) and perpendicular (⊥) to the *c*-axis of the sample between $T_c$ and 25 K. Below $T \approx 25$ K, *T* onset ($T^{on}$) and *T* of zero ($T^{zero}$) resistance are affected by the surface superconductivity, and depend on the applied electrical current density.[23,24] The $H_{c2}$ values obtained from magnetic hysteresis loops (not shown) and the $H_{c2}$ at 20 K obtained from I-V curves are also included on the graph. Extrapolating to *T*=0 K, we obtain $H_{c2}(0)$ of about 3.5 T, which is close to values found in clean single crystals.[25] Using this value, we obtain $\xi_{ab}(0)$ = 10 nm from $H_{c2}^{c} = \Phi_0/[2\pi\xi_{ab}^2(0)]$. Fig. 1(b) shows the results of the $H_{c2}(\theta)$ measurements at 35 K, using the $T^{zero}$ criteria when the voltage drops to zero,[23] and the corresponding fit to the effective mass description $H_{c2}(T,\Theta) = H_{c2}(T,\Theta=0)\varepsilon(\Theta)$, where $\varepsilon(\Theta)=[\cos^2\Theta+\gamma^{-2}\sin^2\Theta]^{1/2}$, where $\Theta$ is the angle between the applied magnetic field **H** and the crystallographic *c*-axis and $\gamma$ is the anisotropy of the critical field. The cusp-like behavior of the experimental data when the field is close to being parallel to the surface (around $\Theta = 90°$) is due to the surface superconductivity.[23] However, considering that surface superconductivity produces a field enhancement of $H_{c3}=1.69H_{c2}$,[26] the $\gamma \approx 2.5$ obtained via the fit) shows a good agreement with anisotropy values obtained in clean single crystals.[23] Analyzing of the data at $T = 24$ K (not shown) with the same $T^{zero}$ criteria, we obtain $\gamma \approx 4.5$. Although the two-band effects result in deviations from the effective mass description, the obtained values of $\gamma$ are in good agreement with those reported in clean systems,[23] when $H_{c3}$ effect is taken in to account.

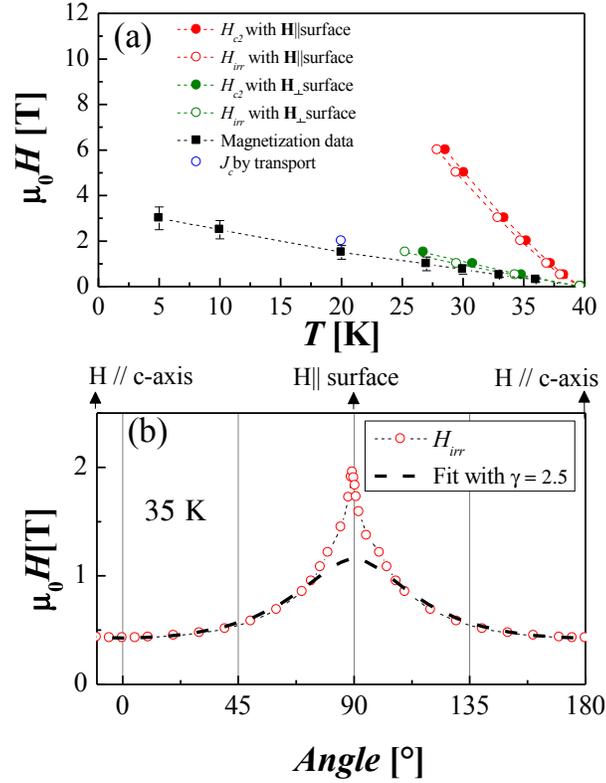

Figure 1. (a) Temperature dependence of the upper critical field ($H_{c2}$) and the irreversibility line ($H_{irr}$) in a MgB$_2$ thin film. (b) Angular dependence of $H_{irr}$ at 35 K and fit by using the anisotropic mass equation described in the text.

Figure 2(a) shows superconducting vortices in the MgB$_2$ film resolved by MFM as bright spots, a result of a repulsive interaction between the tip and the vortices, with magnetization antiparallel to each other. The dark spots in Fig. 2(a) are not anti-vortices since their shapes are irregular compared to the vortices. Instead, they represent nanoscale-size local inhomogeneities. Figures 2(b) and 2(c) show single vortices obtained from the MgB$_2$ and a Nb film reference with the same experimental condition in a single cool-down. Their line profiles along dotted lines in each image are shown in Fig. 2(d). The peak magnitude of MgB$_2$ is larger than that of Nb, indicating the λ value in MgB$_2$ is smaller than the one in Nb.[27] The magnetic penetration depth (λ) at 4 K was estimated by the MFM Meissner method, described elsewhere,[12,27] resulting in λ$_{ab}$=50 ± 10 nm.[28] This value is in agreement with those reported in clean single crystals.[11] Measurements of $J_c$ at 5 K in MgB$_2$ nanobridges of 150 nm in width, obtained from similar quality of thin films, show $J_c \approx 160$ MA cm$^{-2}$.[16] Considering geometry effects,[29] the reported value can be considered as depairing critical current ($J_0$), obtained in a vortex-free state at low temperatures, for π-band contribution.[16,26] At high field the anomalous evolution of ξ(H) and λ(H) predicts low $J_0$ (0 K) value.[3] The theoretical $J_0$ can be estimated via the Ginzburg-Landau equation $J_0^{GL} = \dfrac{cH_c}{3\sqrt{6}\pi\lambda}$, where $c$ is the speed of light in vacuum and $H_c = \dfrac{\Phi_0}{2\sqrt{2}\pi\lambda(0)\xi(0)}$ is the thermodynamic critical field. Using $J_0$=160 MA cm$^{-2}$ and λ$_{ab}$= 50 nm, we obtain ξ$_{ab}$≈25 nm and $H_c \approx$ 1800 Oe. The $H_c$ value is close to those obtained from specific heat measurements, attributed to the π- band in clean single crystals (1500 Oe).[30] In addition, using λ$_{ab}$= 50 nm and ξ$_{ab}$=25 nm, we obtain Ginzburg-Landau parameter $k = 2$, which is within the range of the reported values in clean single crystals at low field.

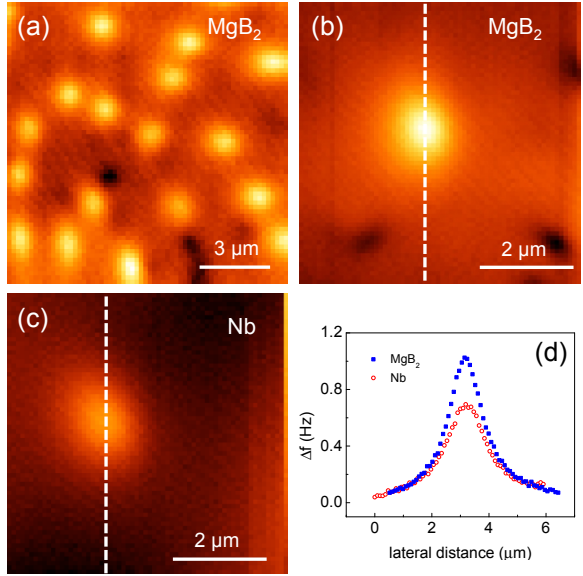

Figure 2. (a) Large field of view image of a MgB$_2$ film. Bright spots are vortices. Dark spots represent local inhomogeneties in the film. (b)-(c) show a single vortex image of MgB$_2$ and Nb, respectively. (d) MFM profiles (shift of the resonance frequency of a magnetic cantilever) of the vortices along the dotted lines in (b) and (c).

Figure 3(a) shows the log-log plot of $J_c$ vs $H$ at four different temperatures (5, 10, 20, and 30 K). The self-field of $J_c$ at $T$=4 K is 26 MA cm$^{-2}$, which is about 16% of the theoretical value of $J_0$ ≈ 160 MA cm$^{-2}$. Three clear distinct regimes can be identified in the log-log $J_c(H)$ plot, similar to those reported on clean MgB$_2$ single crystals.[14] The first regime is between 0 and $B^*$, where $J_c(H) \approx$ *constant*, the second regime is represented by a power-law dependence ($J_c \propto H^\alpha$), and the third depicts a fast drop of the $J_c(H)$. The first regime is clearly seen only at low temperatures. Although this regime has been discussed in cuprates as a single vortex regime,[31] $B^*$ (5 K) ≈ 800 Oe is in good agreement with a crossover produced by self-field effect which can be estimated as $B^*=J_c t$, with t the thickness.[32] The second regime ($J_c \propto H^{-\alpha}$) can be fit with $\alpha$=1. The $\alpha$ value in cuprates and pnictides superconductors is related to the type of pinning centers, and ranges between 0.6 and 0.2 depending on their geometry.[29, 31, 33,34] In cuprates an $\alpha$=1 value is well described by the theory for strong pinning, when the vortex excursion driven by thermal fluctuations is comparable to the inter-vortex distance.[31] However, high pinning energy and low vortex fluctuations in MgB$_2$ require a different analysis.[18] In this context, dirty $MgB_2$ films, presenting lower self-field $J_c$ values at low temperatures but technologically more favorable $J_c$ (*H*) dependences,[18,35] could be analyzed as a high density of strong pinning centers,[36] and show a more isotropic behavior produced by large intra-band scattering.[12] In addition, large intra-band scattering increases λ, and the physics bears similarity to that of single band materials, which reduces $J_0$ and affects the pinning energy as we discuss below. In this context we analyze the pinning landscape in conjunction with fundamental superconducting properties in our film. Figure 3(b) shows the angular dependence of $J_c$ at 35 K at different *H*. The measurements can be understood via anisotropic scaling of the $J_c$ (θ), however, some clear features, associated with different type of strong pinning centers, are evident. The small peak at $\mu_0 H < 0.3$ T when **H**// *c*-axis represents the presence of correlated disorder,[37] whereas the small shoulder at θ ≈ 50 ° (see inset) can be associated with the pinning by small MgO precipitates,[38] as occurs in YBCO films when pinning is dominated by nanoparticles.[39] The peak effect by small nanoparticles is more pronounced at *T* =5 K and high *H* (not shown). In order to understand the pinning nature, we analyzed a field dependence of the pinning force $F_P = J_c H$. When the same type of pinning mechanism dominates over a certain temperature range, the $F_p$ (*H*, *T*) can be scaled as follows: $F_p/F_p^{max} \propto h^m(1-h)^l$, *m* and *l* are exponents that depend on the pinning mechanism, and $h=H/H_{c2}(T)$.[40] Figure 4 shows the $F_p/F_p^{max}$ versus *h* at different temperatures. The curves show a peak at small *h*, which could be associated with $B^*$, and a second broad peak with maximum around *h*≈0.2. This behavior is similar to that found in clean single crystals.[14] In spite of deviations at low field, the curves can be scaled with *m*=0.5 and *l*=2 (see Fig. 4), showing a maximum value around *h*=0.2, which can be associated with normal surface pinning. The fact that normal surface pinning dominates the vortex dynamics in clean $MgB_2$ samples is consistent with the strong suppression of $J_c$ with thickness,[17] and the same mechanism dominates in $MgB_2$ single crystals with lower $J_c$ values. We propose that at low *H* the pinning is strongly dominated by the surface pinning, while, as shown in Fig. 3(b), large defects such as correlated disorder and nanoparticles are more important at high *H* values. Fig. 3 is also consistent with grain boundaries, present at low density, being the source of correlated pinning, which is manifested in suppression of the $J_c$ peak when H//c-axis at $\mu_0 H = 0.3$ T. In this scenario, any evolution of ξ(*H*) and λ (*H*) should modify the pinning energy scales, thus making the relative importance of the pinning mechanisms *H* dependent. Finally, we discuss the third regime associated with a fast drop of $J_c$ (*H*). The value of *H*, where a fast drop starts at each temperature, is similar to those found in single crystals,[14] ruling out thickness/λ ratio as the mechanism responsible for the vortex dynamics.[41] The similitude between $J_c$ (*H*) in our film and those found in single crystals,[14] i.e., a fast drop of $J_c$ around $\mu_0 H = 1$

T at 5 K, suggests that the pinning is affected by intrinsic superconducting properties such as suppression of the π-band, and decrease in the interband coupling.[3,42]

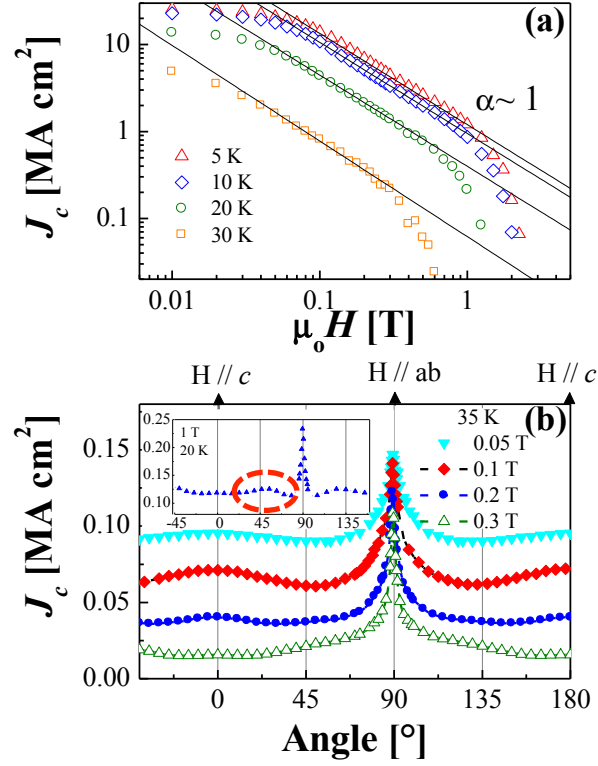

Figure 3. (a) Critical current density ($J_c$) vs. magnetic field ($H$) at different temperatures obtained by using the Bean model. (b) Angular dependence of the critical current density $J_c$ (θ) at 35 K in different applied magnetic fields ($\mu_0 H$ = 0.05, 0.1, 0.2 and 0.3 T). Inset: $J_c$ (θ) at 20 K and $\mu_0 H$ = 1 T. Red circle shows the peak of $J_c$ at θ≈50°.

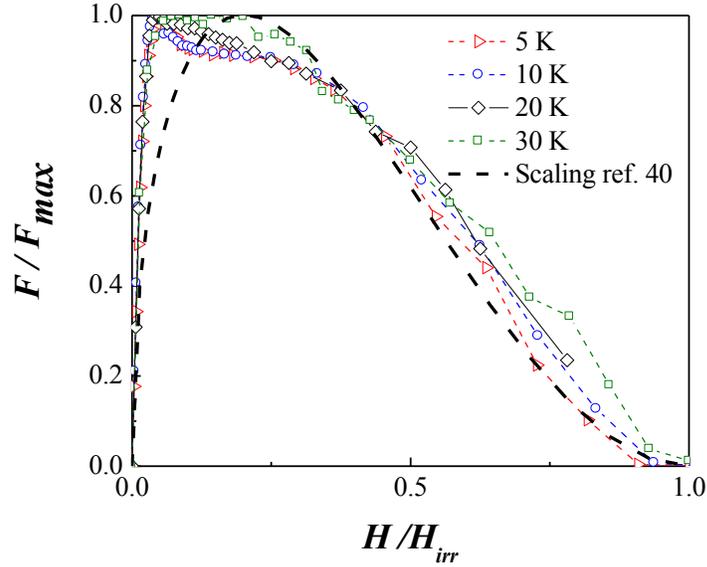

Figure 4. Normalized pinning force ($F_p$) versus normalized magnetic field [$h = H/H_{irr}(T)$] at different temperatures.

Figure 5 shows the temperature dependence of the flux creep rate $S(T)$ for different $H$. The values are very small in comparison with cuprates[43] and pnictides.[44] Similar measurements of vortex dynamics in a MgB$_2$ film were reported by J. Thompson and coworkers (see reference 18). The $J_c(t)$ dependence varies logarithmically according to the simple Anderson–Kim model [45] as $J(t) = J(0)[1 - \left(\frac{T}{U_0}\right)\ln\left(\frac{t}{t_0}\right)]$. The basic concept of the flux creep is that a flux line or a flux bundle can be thermally activated to overcome the pinning energy barrier $U_0$. At low temperature $U_0$ is independent of $T$, and $S$ should change lineally with temperature: $S = T/U_0$. The $U_0$ values, derived from the measured values of $S$ at low temperature, are shown in the inset of Fig. 5. Although our data can be described well with $U_0 = AH^{-0.55}$, the analysis of the $U_0$ in a wide range of $H$ in MgB$_2$ shows a good correlation with $U_0 = AB^\gamma(1 - \frac{B}{B^*})^\delta$.[18] Neglecting the parabolic suppression, our fit results in $A \approx 3600$, which is higher than those reported in dirty films,[18] and indicates that beyond the presence of strong pinning centers, $U_0$ depends on the intrinsic superconducting properties. A rough estimate of the pinning energy, considering the condensation energy within a coherence volume,[19] $U_0 = \left(H_c^2/8\pi\right) {^4\!/_3} \pi \xi^3$, shows that the value is strongly dependent upon both $\lambda$ and $\xi$, which gives rise to $H$ dependence of $J_c$ in clean MgB$_2$.[9] Using our previous estimates of $H_c \approx 1800$

Oe and ξ ≈ 25 nm, we obtain $U_0 \approx 200000$ K, which is in agreement with the expectation at low magnetic field (see inset of Fig 5).

Our results show that clean $MgB_2$ films present two different regimes in $J_c$ ($H$). One is for $\mu_0 H < 1$ T, which is associated with the suppression of the π-band, and the other is for field above 1 T, which is associated with more anisotropic σ-band.[3] Large vortex fluctuations[46] and vortex dissipation[47] have been reported in clean systems at high fields, which is in agreement with a clear change on the vortex dynamics. We found a strong dependence of $U_0$ ($H$), which can be associated with two-band features. The extrapolation of $U_0$($H$) at high magnetic fields suggest very low values which suggest that many of the features of the vortex dynamics in the range dominated by the σ-band in clean $MgB_2$ samples should be similar to those found in pnictides[44] and HTS cuprates.[43] In these systems, the combination of high κ and γ values reduce the pinning energy $U_0$ and the creep rate is higher than in conventional low-$T_c$ superconductors.

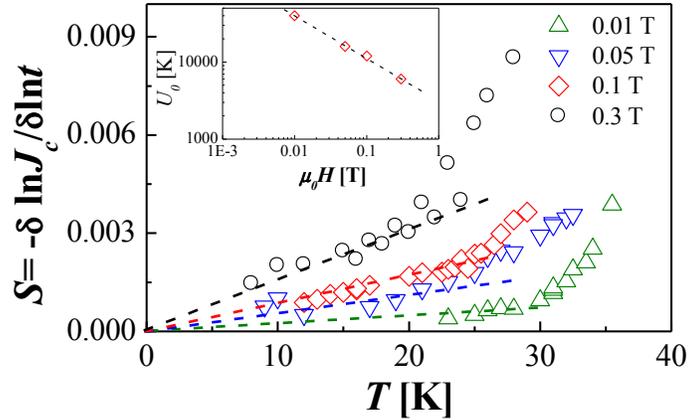

Figure 5. Temperature ($T$) dependences of the creep rate $S = -\dfrac{d(\ln J_c)}{d(\ln t)}$ at different applied magnetic fields. Inset: Pinning energy obtained from $S = T / U_0$ (see the dashed line main panel).

**Conclusions**

We studied intrinsic superconducting properties of a clean $MgB_2$ thin film. We found λ(4 K) = 50 ± 10 nm and $\xi_{ab}$ = 10 nm, which results from the effect of the π-band (low field) and σ-band (high field) on superconductivity. Note sure what you want to say here. The anisotropy of the superconducting critical field $\gamma_{Hc2}$ ranges from 2 close to $T_c$ to 4.5 at low temperature, similar to the values reported previously in clean single crystals. We found high values of self-field $J_c$, that are strongly reduced by $H$, which can be attributed to the suppression of the π-band. The temperature dependence of the creep rate $S$ ($T$) is consistent with a simple Anderson-Kim mechanism. Our findings show that different field dependence of the gaps in a multiband superconductor play an important role in defining its vortex dynamics.